\documentclass[a4paper,paper,notoc]{JHEP3}
\usepackage{amssymb,amsmath}
\usepackage{graphicx}
\usepackage{epsfig}

\newcommand{\beq}{\begin{equation}}
\newcommand{\eeq}{\end{equation}}
\newcommand{\bea}{\begin{eqnarray}}
\newcommand{\eea}{\end{eqnarray}}
\newcommand{\vc}[1]{{\textbf{#1}}}
\newcommand{\mc}[1]{\mathcal{#1}}
\newcommand{\cH}{\mathcal{H}}

\author{Kari Enqvist and Gerasimos Rigopoulos \\
Helsinki Institute of Physics, P.O.Box 64,
FIN-00014, University of Helsinki,
Finland
}

\title{Non-linear mode coupling and the growth of perturbations in $\Lambda$CDM}

\abstract{Cosmic structures at small non-linear scales $k>L\sim 0.2
h $ Mpc$^{-1}$ have an impact on the longer (quasi-)linear wavelengths
with $k<L$ via non-linear UV-IR mode coupling. We evaluate this
effect for a $\Lambda$CDM universe applying the effective fluid method
of Baumann, Nicolis, Senatore and Zaldarriaga \cite{Baumann:2010tm}.
For $k<L$ the $\Lambda$CDM growth function for the density contrast
is found to receive a scale dependent correction and an effective
anisotropic stress sources a shift between the two gravitational
potentials, setting $\phi$ - $\psi \neq 0$. Since such a situation is
generically considered as a signature of modified gravity and/or
dark energy, these effects should be taken into account before any conclusions on the dark sector are drawn from the interpretation of future observations.}

\begin{document}

\section{Introduction}

The standard $\Lambda$CDM model of the
universe is characterized by a specific evolution of
the density contrast $\delta\rho/\rho \equiv \delta \propto a D(a)$ i.e by a specific \emph{growth function} $D(a)$ for the matter perturbations, and by the equality of the two gravitational
potentials $\phi$ and $\psi$ as is implied by the lack of
anisotropic stress in the CDM. 
In contrast, evolving
dark energy, or a modification of Einstein gravity, would generically give rise
to an anisotropic stress as well as to modifications in the Poisson
equation \cite{Amendola:2007rr}, which would manifest themselves in a different functional form for $D(a)$. Probing the growth function for
different values of scale factors $a$ would thus be invaluable to
determine the dark energy related cosmological parameters. It is for
this reason that the three-dimensional mapping of the structure of
the universe by future weak lensing surveys such as EUCLID \cite{euclid} is very
much expected to significantly constrain dark energy and modified
gravity models.

At a practical level the determination of the growth function
proceeds from the assumption that the universe can be considered as
a collection of ideal fluids with some small perturbations. In this setting, an accurate (scale independent) parametrization for D reads \cite{Linder:2005in}
\beq
D(a)=\exp\left[ \int\limits_0^a d\ln{\bar{a}}\left(\Omega_m(\bar{a})^\gamma-1\right)\right]
\eeq
where $\Omega_m({a})=H^2_0\Omega_m a^{-3}/H(a)^2$ and $\gamma$ ($=\frac{6}{11}$ for $\Lambda$CDM) depends e.g. on the equation of state parameter of the dark energy.\footnote{It is interesting to note that scale dependent growth functions are physically more realistic for dark energy models \cite{Dent:2009wi}.} However,
structure formation itself feeds back on the stress-energy and hence
on the effective behavior of the cosmological fluids. As has been
pointed out in \cite{Baumann:2010tm}, this effect can be studied by integrating out the short-wavelength perturbations, obtaining an effective theory for the
long-wavelength universe that
has an equation of state different from the homogeneous background
and, moreover, is no longer strictly ideal but is also characterized
by a viscosity parameter. Very roughtly, one can think that this effect is caused
by the  motion of small-scale lumps of matter and the tidal effects of longer wavelength perturbations on this motion. However, one should also note that the scales that have virialized
can actually be shown to decouple completely from the large-scale dynamics at
all orders in the post-Newtonian expansion, apart of course from contributing a correction to the energy density \cite{Baumann:2010tm}; in effect,
for a distant observer virialized systems behave as if they were point particles.

In the  present paper we follow this approach and study the
effective cosmological fluid of the long-wavelength perturbations in
a pure $\Lambda$CDM universe. We point out that the naive
expectations for e.g. the growth function in the linear regime are violated at some level
due to the effective pressure and anisotropic stress generated from
the non-linear mode coupling between short and long wavelengths.
This effect is generic and scale-dependent and should therefore be
taken into account before any conclusions about the properties of
dark energy are drawn from the interpretation of observations. On
the other hand, it could also be used as a consistency check of the
concordance model.

We proceed with our investigation as follows: in the next section we briefly review the ingredients of the effective fluid approach
to the non-linear coupling of long to short scales \cite{Baumann:2010tm} that we need to derive our results.
Using the methods described in section 2, we compute in section 3 the evolution of perturbations in this effective fluid and the corresponding corrections to the linear $\Lambda$CDM growth function,
as well as the so-called gravitational slip: the divergence of the two gravitational potentials $\phi - \psi$. In the process we also provide an exact analytic expression for the linear $\Lambda$CDM growth function which, to our knowledge, has not appeared in the literature previously. We summarize and conclude in section 4.

\section{The Effective fluid}
As matter perturbations in the universe grow under the influence of
gravity there finally comes a point for a given scale where the
density contrasts become larger than unity. At this point naive
perturbation theory breaks down and other techniques need to be used
in order to follow the further evolution of perturbations,
eventually having to resort to numerical simulations. In our
universe scales with $k>k_{\rm nl}\sim 0.2 h$ Mpc$^{-1}$ have
entered the non-linear regime.

Apart from the difficulty of following the evolution of non-linear
perturbations, another issue is the inevitable coupling between
different scales which is absent in the linear regime. In particular
long and short scales influence each other and it is expected that
the formation of non-linear perturbations might have an impact on
larger, linear scales through UV-IR coupling. It has even been
suggested that even the observed acceleration might be due to such
non-linear effects, and this possibility has recently spurred a
lively debate in the literature \cite{DEdebate}. Regardless of
whether such an explanation for the observed acceleration is
feasible, the UV-IR coupling is always present and should be
quantified, especially given the accuracy of forthcoming
observations which promise to map the universe and its history with
unprecedented accuracy.

The authors of \cite{Baumann:2010tm} address this non-linear mode
coupling in a way that is simple and physically intuitive. By
smoothing out short scale non-linearities they are led to an
effective approach for longer wavelengths where the CDM fluid on
scales $k<k_{\rm nl}$ is replaced by an effective fluid which,
unlike the underlying CDM matter content, possesses anisotropic
stress and pressure. In more detail, the basis of this approach is an effective energy momentum tensor,
including the effects of gravity, which is smoothed out on a scale
$L\leq k_{\rm nl}$. To second order in the derivatives of the
gravitational potential and peculiar velocities, this procedure adds
to the rhs of the Einstein equations the terms
\bea\label{t00}
[\tau^0{}_0]_L &=&-[\rho v_s^iv_s^i]_L-\frac{[\partial_i \phi_s
\partial_i\phi_s ]_L- 4[\phi_s \partial^2\phi_s]_L}{8\pi G a^2}+
O\left[(\partial v_l)^2/L^2 , (\partial\phi_l)^2/L^2\right] \\
\label{tij} [\tau^i{}_j]_L &=&[\rho v_s^iv_s^i]_L-\frac{[\partial_i
\phi_s \partial_i\phi_s ]_L\delta^i_j- 2[\partial_i\phi_s \partial_j
\phi_s]_L}{8\pi G a^2}+ O\left[(\partial v_l)^2/L^2 ,
(\partial\phi_l)^2/L^2\right]\,. \eea The subscripts $l$ and $s$
refer to ``long'' modes with $k<L$ and ``short'' modes with $k>L$
respectively. By the notation $[X]_L$ we mean
\beq\label{average}
[X]_L(\vc{x})\equiv\int d^3\vc{x}'W_L(|\vc{x}-\vc{x}'|)X(\vc{x}')\,,
\eeq
where $W_L(|\vc{x}-\vc{x}'|)$ is a window function that
eliminates modes with $k>L$. Note that since we are only considering
terms linear in metric perturbations (but not their spatial
derivatives) and quadratic in velocities, the expressions in
$\tau_{\mu\nu}$ coincide to this order with proper volume averages.
For $k \ll L$ the third terms on the rhs  of (\ref{t00}) and
(\ref{tij}) are suppressed. Note also that the expressions for
(\ref{t00}) and (\ref{tij}) contain  short wavelength modes with
$k>L$ but the Fourier modes of $[\tau^0{}_0]_L$ and $[\tau^i{}_j]_L$
themselves are only defined for $k<L$. For the discussion that
follows it is the spatial part $[\tau^i{}_j]_L$ that will be most
important.

According to \cite{Baumann:2010tm}, the effective stress tensor can
be expressed as \beq [\tau^i{}_{j}]_L(\vc{x}) =
\langle[\tau^i{}_{j}]_L\rangle + \Delta\tau^i{}_{j}(\vc{x}) +
\alpha^i{}_{j}(\vc{x})\,. \eeq with the first and second terms
admitting an effective description to lowest order in $(k/k_{NL})^2$
as the stress tensor of an \emph{imperfect} fluid with pressure and
anisotropic stress \beq
\langle[\tau^i{}_{j}]_L\rangle+\Delta\tau^i{}_{j}=\left(\mc{P}+\Delta\mc{P}\right)\delta^{i}_j+\Sigma^i{}_j\,,
\eeq with $\Sigma^i{}_i=0$. In particular, the pressure terms can be
written as \beq \mc{P}\equiv \frac{1}{3}
\langle[\tau^i{}_{i}]_L\rangle= w \rho\,, \eeq \beq\label{deltaP}
\Delta\mc{P}=c_s^2 \rho \,\delta \eeq where $w$ and $c_s^2$ are the
equation of state parameter and the sound speed squared, while the
scalar anisotropic stress is given as \beq\label{sigma}
\Sigma^i{}_j=\eta\left(\frac{k_{i}k_j}{k^2}-\frac{1}{3}\delta^i_j\right)k^2v~,
\eeq where $\eta>0$ is the coefficient of shear viscosity. Note that
\beq \langle[\hat{\tau}^i{}_{j}]_L\rangle=0~, \eeq where $
\hat\tau^i{}_j=\tau^i{}_j-\delta^i_j/3\,\tau^l{}_l$, so there is no
zero mode for the anisotropic stress. It is important to stress at
this point that although the universe only contains CDM, the
effective description on long wavelengths involves both pressure and
shear viscosity. It is customary to define the anisotropy scalar
$\sigma$ as \beq
\sigma\equiv-\frac{1}{\rho}\frac{k_ik_j}{k^2}\Sigma^i_j =
-\frac{2}{3}\frac{\eta}{\rho}k^2v\equiv c_{vis}^2\frac{-k^2v}{\cH}
\eeq where we have defined the dimensionless parameter
$c_{vis}^2\equiv 2 \eta\cH/3\rho$.

The possibility of expressing the effective stress-energy tensor in
terms of long wavelength perturbations such as in eqs (\ref{deltaP})
and (\ref{sigma}) is a manifestation of non-linear mode coupling
relating the short wavelength modes in the definition of
$[{\tau}^i{}_{j}]_L$ to modes with $k<L$. The effective fluid is
then described by two parameters $c_{s}^2$ and $c_{vis}^2$ which can
be determined by using the definition (\ref{tij}) via
\bea\label{param1} c_{s}^2 &=&
\frac{1}{3\rho}\frac{\langle[\tau^i{}_i]_L\,\delta_l\rangle}{\langle\delta_l\,\delta_l\rangle}\\
\label{param2} c_{vis}^2
&=&-\frac{\cH}{\rho}\frac{k_ik_j}{k^2}
\frac{\langle[\hat\tau^i{}_j]_L\,\nabla\cdot\vc{u}_l\rangle}{\langle\nabla\cdot\vc{u}_l\,\nabla\cdot\vc{u}_l\rangle}\,.
\eea The tidal effects of long wavelength gravitational
perturbations on shorter wavelength modes produce correlations of
the long wavelength $\delta$ and $v$ to the short wavelength modes
in $[{\tau}^i{}_{j}]_L$, making $c_{s}^2$ and $c_{vis}^2$ different
from zero.

Finally, there will also be a stochastic part, given by the
$\alpha^i{}_{j}$ term.  It encodes deviations from the average value
that are uncorrelated to long wavelength variables and acts as an
external source. The variance of these terms can be obtained via
\beq\label{stochastic}
\langle\alpha^i{}_j\,\alpha^k{}_l\rangle=\langle[\tau^i{}_j]_L[\tau^k{}_l]_L\rangle
-\langle[\tau^i{}_j]_L\rangle\langle[\tau^k{}_l]_L\rangle-\langle\Delta\tau^i{}_j\Delta\tau^k{}_l\rangle
\,. \eeq

Summarizing, we see that the UV-IR coupling can be described by the use of an effective theory where the short
scale fluctuations act as an effective \emph{imperfect} fluid on longer wavelengths. The Einstein
linearized equations with $\Delta\tau^i{}_j+\alpha^i{}_{j}$ added to the rhs can then be used to follow the
linear response of long wavelength perturbations to short wavelength non-linearities. The effective fluid is
characterized by two parameters: $c_{s}^2$ and $c_{vis}^2$. The correlators on the rhs of (\ref{param1}) and (\ref{param2})
can be determined from small-scale N-body simulations using the definitions (\ref{t00})
and (\ref{tij}) for $[\tau^0{}_0]_L$ and $[\tau^i{}_j]_L$.
Once these parameters are determined, no further reference to the
short scale dynamics is needed. Alternatively, the parameters of the
effective fluid could be considered as free parameters to be fitted
from observations - see section 3.

\section{Perturbations of the effective fluid in $\Lambda$CDM}

\subsection{Density contrast and gravitational slip}

We can now proceed to evaluate the effect of short wavelength
fluctuations on longer wavelengths using the approach outlined
above. We start with the Einstein equations relating the
gravitational potentials with the components of the effective energy
momentum tensor. The $00$ and $0i$ equations give \beq k^2\phi
+3\cH\left(\dot{\phi}+\cH\psi\right)=
-\frac{3}{2}H_0^2\frac{\Omega_m}{a}\delta \eeq \beq
\dot{\phi}+\cH\psi=-\frac{3}{2}H_0^2\frac{\Omega_m}{a}(1+w)v~, \eeq
while the traceless part of the $ij$ equation gives \beq
k^2(\phi-\psi)=\frac{9}{2}H_0^2\frac{\Omega_m}{a}(1+w)\sigma\,. \eeq
We can further use the energy-momentum, conservation law $\nabla_\mu
T^{\mu\nu}=0$ which results in \beq
\dot{\delta}=(1+w)\left(k^2v+3\dot{\phi}\right)-3\cH\left(\frac{\delta
\mc{P}}{\rho_m}-w\delta\right)~, \eeq \beq
\dot{v}=-\left(1-3w+\frac{\dot{w}}{1+w}\right)\cH v
-\frac{1}{1+w}\frac{\delta \mc{P}}{\rho_m}+\sigma-\psi \eeq with $v$
the peculiar velocity potential: $u_i=ik_iv$.

According to the discussion in the previous section we can express
the pressure perturbation and the anisotropic  stress of the
effective fluid in terms of long wavelength perturbations $\delta$
and $v$ and a stochastic part uncorrelated with them \beq \delta
\mc{P}=c_s^2\rho_m\delta + c_s^2\rho_m\alpha_1 \eeq \beq
\sigma=c_{vis}^2\frac{-k^2v}{\cH}+c_{vis}^2\alpha_2 \simeq
c_{vis}^2\tilde{\delta}+c_{vis}^2\alpha_2\,, \eeq where we have used
the linear perturbation theory relation
$\delta\simeq-k^2v/\cH$ for the (long wavelength) velocity.
The quantities $\alpha_1$ and $\alpha_2$ encode the (dimensionless)
stochastic fluctuations of $\delta P$ and $\sigma$. We stress that
we ignore here the non-linearities of the long wavelength
perturbations\footnote{We could have included terms quadratic in the
gradients of the potentials and the velocities, studying
non-linearities on long wavelengths via second order perturbation
theory}.

We can now obtain an equation for the evolution of the long wavelength density contrast $\delta$. To simplify the result we make the following assumptions:
\begin{enumerate}
  \item We are interested in following perturbations on scales smaller than the horizon so that $k^2 > \cH^2$.
  \item  We will also assume that $c_s^2$ and $c_{vis}^2$ are small enough such that $c^2\frac{k^2}{\cH^2} < 1$.  This requires $c^2<10^{-5}\left(\frac{k_{\rm nl}}{L}\right)^2$ where $L$ is the smoothing scale we use to define the long wavelength sector and $k_{\rm nl}\sim 0.2$ h Mpc$^{-1}$ is the scale of non-linearity.
  \item The time scales of evolution on scales $k>L$ is assumed not to be much faster than cosmological time scales, ie $\frac{d}{d \eta}(c \alpha) \sim \cH c \alpha$ and that
  \item the timescale for evolution for $w$ is similar $\frac{d w}{d\eta}\sim \cH w$.
\end{enumerate}
Assumptions 2), 3) and 4) can of course be checked once the parameters have been calculated from first principles.

Given the above we obtain for the density perturbation to leading
order in $c^2$
\beq \ddot{\delta} + \cH\dot\delta -
\frac{3}{2}H_0^2\frac{\Omega_m}{a}\delta=-k^2c_s^2\delta
-k^2\left(c_s^2\alpha_1-c_{vis}^2\alpha_2\right)~, \eeq
which
reduces to the standard equation when short-scale non-linearities
are ignored ie $c_s^2\rightarrow 0$ and $c_{vis}^2\rightarrow 0$. It
is useful to express the above in terms of the scale factor $a$.
Using $\frac{d}{d\eta} = a\cH \frac{d}{da}$ we have for $\delta(a)$
\beq \frac{d^2}{da^2}\delta+(\frac{d}{da}\ln \cH
+\frac{2}{a})\frac{d}{da}\delta -
\frac{3}{2}\left(\frac{H_0}{\cH}\right)^2\frac{\Omega_m}{a^3}\delta
= -\frac{1}{a^2}\frac{k^2}{\cH^2}\left(c_s^2\delta
+c_s^2\alpha_1-c_{vis}^2\alpha_2\right)~. \eeq
We denote the solution
in the absence of mode coupling (rhs is taken to be zero) by
$\tilde{\delta}$ \beq\label{delta-eq-hom}
\frac{d^2}{da^2}\tilde\delta+(\frac{d}{da}\ln \cH
+\frac{2}{a})\frac{d}{da}\tilde\delta -
\frac{3}{2}\left(\frac{H_0}{\cH}\right)^2\frac{\Omega_m}{a^3}\tilde\delta
= 0~. \eeq Then, to leading order in $c^2$, the solution of the long
wavelength density contrast in the presence of  small scale
non-linearities is
\beq\label{delta-Green-solution}
\delta(a)=\tilde\delta(a) -\int\limits_{a_{in}}^\infty dx
\,G(a,x)\frac{1}{x^2}\frac{k^2}{\cH(x)^2}\left(c_s^2\tilde\delta(x)
+c_s^2\alpha_1-c_{vis}^2\alpha_2\right)~, \eeq
where $G(a,x)$ is the
appropriate Green function satisfying
\beq
\frac{d^2}{da^2}G(a,x)+(\frac{d}{da}\ln \cH
+\frac{2}{a})\frac{d}{da}G(a,x) -
\frac{3}{2}\left(\frac{H_0}{\cH}\right)^2\frac{\Omega_m}{a^3}G(a,x)
= \delta_D(a-x)\,. \eeq

The solution to (\ref{delta-eq-hom}) that grows like $\tilde\delta
\propto a$ at early times (during matter domination)
reads
\beq\label{delta-solution}
\tilde\delta(a)=\frac{\tilde\delta_{in}}{a_{in}}
\frac{5}{2}H_0^2\Omega_m\frac{\cH(a)}{a}\int\limits_{0}^a\frac{dx}{\cH(x)^3}\,,
\eeq
where $a_{in}$ is the scale factor at which we start the
computation and $\tilde\delta_{in}$ is the  density contrast at that
time. In a $\Lambda$CDM universe we have
\beq \cH=H_0 \,a
\left(\frac{\Omega_m}{a^3}+\Omega_\Lambda\right)^{1/2}\,, \eeq
and
the integral can be performed to give
\beq\label{integral}
\int\limits^a_0\frac{du}{u^3\left(\frac{\Omega_m}{u^3}+\Omega_\Lambda\right)^{3/2}}=\frac{2}{5\Omega_m^{3/2}}\,a^{5/2}\,
F(a)~, \eeq
where we have defined $F$
\beq
F(a)\equiv\,_2F_1\left(\frac{3}{2},\frac{5}{6};\frac{11}{6};-\frac{\Omega_\Lambda}{\Omega_m}a^3\right)
\eeq
with $_2F_1$ the hypergeometric function. Thus,
\beq
\tilde\delta(a)=\frac{\tilde\delta_{in}}{a_{in}}
\left(1+\frac{\Omega_\Lambda}{\Omega_m}a^3\right)^{1/2}a F(a), \eeq
and the $\Lambda$CDM {growth function} for the density contrast
is \beq\label{D}
D(a)=\left(1+\frac{\Omega_\Lambda}{\Omega_m}a^3\right)^{1/2} F(a)~.
\eeq
The Green function for the problem (satisfying homogeneous
initial conditions at $a=0$) is found to be
\beq
G(a,x)=\Theta(a-x)\frac{2}{5H_0^3\Omega_m^{3/2}}\,\,x^2\cH(x)^2\frac{\cH(a)}{a}\left(a^{5/2}F(a)-x^{5/2}F(x)\right)\,.
\eeq
\begin{figure}[h]
 \centering
 \includegraphics[scale=0.8]{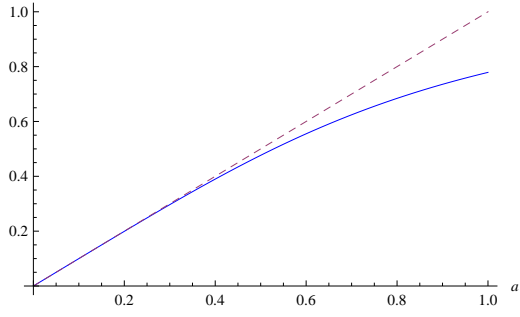}
 \label{LCDM-Growth}
 \caption{$\Lambda$CDM linear growth for density contrast. Dashed line is for the Einstein-de Sitter Universe: $\delta\propto a$.}
 \end{figure}

Using the above, we can now obtain for the solution for the density contrast from (\ref{delta-Green-solution})
\beq\label{density-contrast}
\delta(a)=\left(1-\frac{k^2c_s^2}{H_0^2}E_1(a)\right)\tilde{\delta}(a)-\frac{k^2c_s^2}{H_0^2}\mc{S}(a,\alpha)~,
\eeq
where
\beq
E_1(a)=\frac{2}{5\Omega_m}\int\limits_0^a dx\,\left(1+\frac{\Omega_\Lambda}{\Omega_m}x^3\right)^{1/2}x F(x)\left(1-\left(\frac{x}{a}\right)^{5/2}\frac{F(x)}{F(a)}\right)\frac{c_s^2(a)}{c_s^2}
\eeq
and
\beq
\mc{S}(a,\alpha)=\frac{2}{5\Omega_m}\,a^2F(a)\left(1+\frac{\Omega_\Lambda}{\Omega_m}a^3\right)^{1/2}\int\limits_0^a dx\left(1-\left(\frac{x}{a}\right)^{5/2}\frac{F(x)}{F(a)}\right)\frac{c_s^2(a)}{c_s^2}\alpha\,,
\eeq
with $\alpha \equiv {\alpha}_1-\frac{c_s^2}{c_{vis}^2}\alpha_2$ and $c_s^2$ the speed of sound today.
For the difference of the two gravitational potentials we get
\beq\label{slip}
-k^2(\phi-\psi)=\frac{45}{4}H_0^2c_{vis}^2
\frac{\Omega_m\left(1-\frac{3}{5}\left(1+\frac{\Omega_\Lambda}{\Omega_m}a^3\right)^{1/2}F(a)\right)}
{\left(1+\frac{\Omega_\Lambda}{\Omega_m}a^3\right)^{3/2}a F(a)}\tilde{\delta}(a)
+\frac{9}{2}H_0^2\frac{\Omega_m}{a}c_{vis}^2{\alpha}_2~.
\eeq

\subsection{The matter powerspectrum and weak lensing}\label{impact-on-obs}

In the previous subsection we saw how the impact of UV-IR coupling on the growth of perturbation can be described
in terms of an effective quasi-linear theory on scales larger than
$L$. The main results where equations (\ref{density-contrast}) for
the density contrast and (\ref{slip}) for the gravitational slip. From (\ref{density-contrast}) we
immediately obtain
\beq\label{delta-power}
\langle\delta_\vc{k}\delta_\vc{p}\rangle=\left(1-2\frac{k^2c_s^2}{H_0^2}E_1(a)\right)\langle\tilde{\delta}_\vc{k}\tilde{\delta}_\vc{p}\rangle
-\frac{k^2c_s^2}{H_0^2}\frac{p^2c_s^2}{H_0^2}\langle\mc{S}_\vc{k}\mc{S}_\vc{p}\rangle~,
\eeq
were we have used
$-k^2\phi\simeq\frac{3}{2}\frac{\Omega_m}{a}H_0^2\tilde\delta$. To
simplify our formulae we will make the assumption that
\beq\langle\mc{S}_{\vc{k}}\mc{S}_{\vc{p}}\rangle =
(2\pi)^3\delta(\vc{k}+\vc{p}) \, \beta(a,k)\,P_\delta(k)~, \eeq
where
$P_\delta(k)$ is the density contrast power-spectrum:
$\langle\tilde{\delta}_\vc{k}\tilde{\delta}_\vc{p}\rangle=(2\pi)^3\delta(\vc{k}+\vc{p})
\,P_\delta(k)$. In particular this entails that different modes of
the stochastic fields are \emph{uncorrelated} for different momentum
magnitudes, at least to this order in $c^2$. Given this assumption we can
write \beq
\langle\delta_\vc{k}\delta_\vc{p}\rangle=\left(1-2\frac{k^2c_s^2}{H_0^2}E_1(a)
-\frac{k^4c_s^4}{H_0^4}\beta(a,k)
\right)\langle\tilde{\delta}_\vc{k}\tilde{\delta}_\vc{p}\rangle~, \eeq
where \beq \beta(a,k)\equiv
\frac{\langle\mc{S}_\vc{k}\mc{S}_{-\vc{k}}\rangle}{\langle\tilde\delta_\vc{k}\tilde\delta_{-\vc{k}}\rangle}
\eeq denotes the ratio of the stochastic to the density
power-spectra. Using
\beq
c_s^4\rho_m^2\langle\alpha_\vc{k}\alpha_{-\vc{k}}\rangle\simeq \frac{1}{9}\left(\langle\left[\tau^i_i\right]_{L\vc{k}}^2\rangle-\langle\left[\tau^i_i\right]_{L\vc{k}}\rangle^2\right)
\eeq
we obtain
\beq\label{beta}
c_s^4 \beta \simeq E_2(a)\frac{k_{\rm eq}}{k T(k)^2}\gamma_2
\eeq
where
\beq
E_2(a)=a^4 F(1)^2\int\limits_0^a dx\left(1-\left(\frac{x}{a}\right)^{5/2}\frac{F(x)}{F(a)}\right)^2  \frac{c_s^4(a)}{c_s^4}\,,
\eeq
\beq
\gamma_2 \simeq 1.4\times 10^{-3}\int\limits_L\,\frac{dq}{k_{\rm eq}q^2}\,\frac{H_0^8 P_\delta(q)^2}{\delta_H^2}\,,
\eeq
and $T(k)$ is the transfer function for $\Lambda$CDM (see eg.
\cite{Weinberg:2008zzc} eq. 6.5.12). To obtain (\ref{beta}) we used the fact that since $k$ is in the linear regime $P_{\rm lin}(k)=2\pi^2\delta_H^2\frac{k}{H_0^4}T(k)^2\frac{D(a)^2}{D(1)^2}$ and $\delta_H$ is the perturbation amplitude at the current horizon scale: $\delta_H=4.9\times 10^{-5}$ \cite{Komatsu:2010fb}.

We have now reached our main conclusion. We see that we obtain a \emph{scale dependent} correction to the pure $\Lambda$CDM growth function
\beq\label{Q} D(a) \rightarrow Q(k,a)^{1/2}D(a)\equiv
\left(1-2\frac{k^2}{H_0^2}\,c_s^2E_1(a)
-\frac{k^3}{T(k)^2 H_0^3}\frac{k_{\rm eq}}{H_0} \,\gamma_2E_2(a)\right)^{1/2}D(a)~.
\eeq

This may have interesting ramifications for weak lensing
observations, which probe the potential
$k_ik_jp_lp_m\langle(\phi+\psi)_\vc{k}(\phi+\psi)_\vc{p}\rangle$ \cite{Dodelson:2003ft}.
Because of the UV-IR coupling, this will be also modified from its
pure $\Lambda$CDM form. Defining $\psi=\eta\phi+\lambda$ we obtain
from (\ref{slip}) \beq\label{eta}
\eta=-\frac{15}{2}c_{vis}^2\frac{1-\frac{3}{5}\left(1+\frac{\Omega_{\Lambda}}{\Omega_m}a^3\right)^{1/2}F(a)}
{\left(1+\frac{\Omega_{\Lambda}}{\Omega_m}a^3\right)^{3/2}F(a)} \eeq
and \beq\label{lambda}
\lambda=\frac{9}{2}\frac{\Omega_m}{a}\frac{H_0^2}{k^2}c_{vis}^2\alpha_2(k)\,.
\eeq The weak lensing potential can then be calculated to be \bea
k_ik_jp_lp_m\langle(\phi+\psi)_\vc{k}(\phi+\psi)_\vc{p}\rangle&=&\left[4(1+\frac{\eta}{2})Q(k,a)
+c_{vis}^4\beta_2(k)\right]\nonumber\\
&&\times\left(\frac{3}{2}H_0^2\frac{\Omega_m}{a}\right)^2\frac{k_ik_j}{k^2}\frac{p_lp_m}{p^2}
\langle\tilde{\delta}_\vc{k}\tilde{\delta}_\vc{p}\rangle~, \eea
with $\beta_2(k) \equiv \langle\alpha_{2\vc{k}}\alpha_{2-\vc{k}}\rangle/{\langle\tilde\delta_\vc{k}\tilde\delta_{-\vc{k}}\rangle}\sim \gamma_2(k)$.
We can ignore terms that are not enhanced by factors of $k$, obtaining \beq\label{lensing-potential}
k_ik_jp_lp_m\langle(\phi+\psi)_\vc{k}(\phi+\psi)_\vc{p}\rangle \simeq 4Q(k,a)
\left(\frac{3}{2}H_0^2\frac{\Omega_m}{a}\right)^2\frac{k_ik_j}{k^2}\frac{p_lp_m}{p^2}
\langle\tilde{\delta}_\vc{k}\tilde{\delta}_\vc{p}\rangle~. \eeq
In our case the weak lensing potential directly probes the modified growth function with contributions from the effective anisotropic stress being subdominant.

An accurate determination of the magnitude of the parameters $c_s^2$ and $\gamma_2$ controlling the correction $Q-1$ is beyond the scope of this paper since it would require calculations in the non-linear regime in two distinct ways: 1) Determination of $\gamma_2$ requires knowledge of the non-linear powerspectrum. 2) The effective sound speed $c_s^2$ is a distinctly non-linear phenomenon which at lowest order depends on the three-mode coupling between one long and two short-wavelength modes - see (\ref{param1}) which is zero in linear theory. Therefore, it is more straightforward to consider that a scale-dependent correction to the $\Lambda$CDM growth function due to non-linear mode coupling would have the form
\beq\label{Q-fit} Q(k)=
\left(1-2\frac{k^2}{H_0^2}\,\mu_1
-\frac{k^3}{T(k)^2 H_0^3}\frac{k_{\rm eq}}{H_0} \,\mu_2\right)~,
\eeq
where $\mu_1$ and $\mu_2$ are parameters to be fitted by observation.

However, before closing this section let us make the following observations. We expect that as non-linearities grow, so will the sound speed $c_s^2$. We can thus obtain an upper bound on the value of the functions $E1(a)$ and $E2(a)$ today ($a=1$) by setting $c_s^2(a)/c_s^2\rightarrow 1$. This gives
\beq
E1(a=1)\leq 0.26\,,\quad E2(a=1)\leq 0.1
\eeq
Furthermore, simply extrapolating the linear theory powerspectrum into the non-linear regime to calculate $\gamma_2$ and using 2nd order perturbation theory \cite{Bernardeau:2001qr} to evaluate $c_s^2$ we obtain
\beq
c_s^2\sim\frac{\delta_H^2}{2}\frac{k_{\rm eq}^2}{H_0^2}\int\limits_{\sqrt{2}k_{\rm nl}/k_{\rm eq}}\!\!\!\!\!\!d\kappa \, \kappa \,
T^2(\kappa)\simeq \mc{O}(1) \times 4.2\times 10^{-6}\,,
\eeq
\beq
\gamma_2 \sim \delta_H^2 \!\!\!\!\int\limits_{\sqrt{2}k_{\rm nl}/k_{\rm eq}}\!\!\!\!\!\!d\kappa \,
T^4(\kappa)\sim 10^{-13}
\eeq
We thus see that at the non-linear scale $k_{\rm nl}\sim 0.2$ h Mpc$^{-1}$ linear theory extrapolations predict corrections to the growth function of the order of $10 \%$ and $1 \%$ respectively for the two correction terms in $Q$ (\ref{Q}), with the second term growing faster than the first with k. We stress again that these estimates are based on linear and second order extrapolations in the highly non-linear regime. However, they could be considered indicative of a possibly measurable effect towards the end of the linear regime, accessible through forthcoming combinations of weak lensing and deep redshift surveys such as EUCLID \cite{euclid}.

\section{Summary and Discussion}

We have considered the influence of short wavelength matter fluctuations on
longer wavelengths through non-linear mode coupling in $\Lambda$CDM,
applying the method of \cite{Baumann:2010tm} in this case. Integrating out the short scales gives rise to an effective long-wavelength fluid coupled to gravity that has slightly different properties as compared to the pure $\Lambda$CDM. The effective fluid is viscous and exhibits a small pressure as well as an anisotropic stress, all of which are in part correlated with the longer wavelength perturbations and in part purely stochastic.

We have discussed the nature of these corrections and derived their contributions to the growth function and the weak lensing potential of standard $\Lambda$CDM cosmology. In the process we also provided an analytic expression for the growth function of pure $\Lambda$CDM which, to our knowledge has not appeared  in the literature before, equation (\ref{D}). The contribution from the induced anisotropic stress turns out to be negligible for present observational purposes, but the modification of the growth
function $Q$, which is scale-dependent, could be of importance. It contains a term that scales as $k^2$ and is due
to the perturbation of the effective pressure, while a $k^3/T(k)^2$
term appears because of the stochastic nature of small scale fluctuations
that are uncorrelated to long wavelength variables and effectively act as an
external source in the evolution equation of the long-wavelength perturbation.
The corrections are characterized by two
parameters, $c_s^2$ and $\gamma_2$, which can in principle be
calculated but, perhaps more realistically, can also be fitted from observations. We extrapolated perturbation theory to obtain estimates for these parameters which should more appropriately be considered as lower limits; a more precise estimation would require one to go beyond perturbation theory into the non-linear regime of structure formation.

We would like to end by noting that searches for modifications of
gravity and/or dynamical dark energy focus on the the modifications of the growth
function compared to its $\Lambda$CDM form as well as a non-zero difference in the two gravitational potentials. However, as we have discussed here, such modifications arise at some level even in standard $\Lambda$CDM because of the non-linearity of gravity which couples different scales. In this case, the difference in the two potentials seems too small to be observationally relevant but scale dependent corrections to the growth function might be detectable. These effects are always present and should be taken into account in future surveys before any conclusions are drawn with pertaining to the dark sector. Furthermore, such effects could be used as consistency checks of the standard cosmological model, provided that a more accurate determination of the parameters of the effective fluid is achieved.

\acknowledgments
This research is supported by the
European Union through Marie Curie Research and Training Network
``UNIVERSENET'' (MRTN-CT-2006-035863). KE is also supported by the Academy of Finland
grants 218322 and 131454.

\end{document}